\documentclass{article}

\usepackage{arxiv}
\usepackage[utf8]{inputenc}
\usepackage[T1]{fontenc}    
\usepackage{lmodern}        
\usepackage{hyperref}       
\usepackage{url}            
\usepackage{booktabs}       
\usepackage{amsfonts}       
\usepackage{amsmath,amssymb}
\usepackage{graphicx}
\usepackage{longtable}
\usepackage{xcolor}
\providecommand{\tightlist}{%
  \setlength{\itemsep}{0pt}\setlength{\parskip}{0pt}}
\usepackage{array}
\usepackage{calc} 

\usepackage{float}
\usepackage{caption}
\title{Human-AI collaboration or obedient and often clueless AI in instruct, serve, repeat dynamics?}

\author{
 Mohammed Saqr \\
  University of Eastern Finland\\
  Joensuu, 80130, Finland \\
  \texttt{mohammed.saqr@uef.fi} \\
   \And
 Kamila Misiejuk \\
  FernUniversität in Hagen\\
  Hagen, 58097, Germany \\
  \texttt{kamila.misiejuk@fernuni-hagen.de} \\
  \And
Sonsoles López-Pernas \\
  University of Eastern Finland\\
  Joensuu, 80130, Finland \\
  \texttt{sonsoles.lopez@uef.fi} \\
}

\begin{document}
\date{s }
\maketitle
\date{ s}
\begin{abstract}
While research on human-AI collaboration exists, it mainly examined
language learning and used traditional counting methods with little
attention to evolution and dynamics of collaboration on cognitively
demanding tasks. This study examines human-AI interactions while solving
a complex problem. Student-AI interactions were qualitatively coded and
analyzed with transition network analysis, sequence analysis and partial
correlation networks as well as comparison of frequencies using
chi-square and Person-residual shaded Mosaic plots to map interaction
patterns, their evolution, and their relationship to problem complexity
and student performance. Findings reveal a dominant Instructive pattern
with interactions characterized by iterative ordering rather than
collaborative negotiation. Oftentimes, students engaged in long threads
that showed misalignment between their prompts and AI output that
exemplified a lack of synergy that challenges the prevailing assumptions
about LLMs as collaborative partners. We also found no significant
correlations between assignment complexity, prompt length, and student
grades suggesting a lack of cognitive depth, or effect of problem
difficulty. Our study indicates that the current LLMs, optimized for
instruction-following rather than cognitive partnership, compound their
capability to act as cognitively stimulating or aligned collaborators.
Implications for designing AI systems that prioritize cognitive
alignment and collaboration are discussed.
\end{abstract}


Human interactions with technology have been an important field of
inquiry whether technology has been the medium or the target of
interactions (Crandall et al., 2018). The recent emergence of generative
AI (GenAI) has created a new reality, where GenAI could replicate high
cognitive functions that were hitherto exclusive to humans for the first
time in history (Mustafa et al., 2024). GenAI can maintain a
conversation that may be indistinguishable from humans, generate
high-quality text, or produce sophisticated code or highly detailed text
(Ray, 2023; Anonymized, 2024). These new capabilities have transformed
many of our routine tasks and opened new avenues for several others to
be changed. Learning is not an exception. Pre-GenAI research indicated a
possible positive impact of educational AI tools, particularly chatbots,
on learning outcomes (Anonymized, 2025). Consequently, the rise of GenAI
has sparked significant optimism about its ability to transform and
improve education.

GenAI may be poised to change how students learn, access knowledge, or
interact with learning resources, to mention a few (Järvelä et al.,
2025; Mustafa et al., 2024). Furthermore, GenAI has redefined the
dynamics of human-AI interaction, creating a paradigm shift where
technology is no longer just a tool but may act as an active
collaborator, teacher, or partner who mimics humans in maintaining a
dialogue, responding with empathy, and performing tasks. Notwithstanding
the debate of how capable GenAI is of actual reasoning or understanding,
it maintains a conversation with perceived sympathy and intelligence
enough to engage users in productive dialogue and most probably act as a
collaborator (Järvelä et al., 2025; Li et al., 2024; Pedreschi et al.,
2025; Xu et al., 2022).

These new human-AI collaboration realities have created opportunities
and challenges as well as a need to study the dynamics of how
collaboration happens, evolves, or unfolds throughout time (Sun, 2024).
While research has started exploring these venues, it comes mostly from
language learning and academic writing (Li et al., 2024; Sun, 2024). As
we currently stand, research that explores how students collaborate with
AI to solve complex technical problems is rather rare. A gap that our
study seeks to bridge. In particular, our study aims to investigate how
students use AI to co-create complex networks with multidimensional
characteristics that involve contextual, structural, as well as
statistical properties. This complexity requires students to specify,
verify, and negotiate the outcome to ensure it matches the
specification.

To that end, the aim of this study is to answer the following research
question:

\begin{itemize}
\item
  RQ1: What are the types of interactions that students use to interact
  with AI to solve complex problems?
\item
  RQ2: How do these interactions unfold in time in both sequence and
  dynamics?
\item
  RQ3: How do students' interaction patterns vary according to
  performance level and problem complexity?
\end{itemize}

\section{\texorpdfstring{\textbf{Background}}{Background}}\label{background}

\subsection{The spectrum of human-AI
interactions}\label{the-spectrum-of-human-ai-interactions}

Human-AI interactions ---and machines at large--- have evolved into
diverse forms, spanning a wide spectrum from augmentation to dependency
and from the evolution of human capabilities to devolution and
delegation of duties and responsibilities (Akata et al., 2020; Pedreschi
et al., 2025). Possibly the most common and widely recognized role of AI
systems is assisting humans in performing tasks, e.g., spell-checking
and grammar correction. Long before the recent rise in GenAI, AI systems
have been used to support writing across academic disciplines, offering
a range of functions from grammar and spelling suggestions to more
detailed feedback (Chen et al., 2022; Crompton \& Burke, 2023). In this
case, AI ---though not more than a ``savant'' with a limited and narrow
scope--- would help augment human intelligence and refine their tasks
(Sun, 2024). This combination of human and machine intelligence enhances
human capabilities, enabling meaningful decisions and actions beyond the
reach of humans or machines alone (Akata et al., 2020; Pedreschi et al.,
2025).

As AI advanced, so did the tasks it could perform and the role it was
asked to play. This growth in capabilities marked a shift to more AI
involvement and collaboration where AI evolved from competing at
zero-sum tasks, such as chess or Go, to cooperating or supporting humans
to perform their work (Crandall et al., 2018). Increasingly, AI has
become a collaborator, co-designer, and co-creator, emphasizing
partnership and shared creativity over competition (Hemmer et al., 2021;
Pedreschi et al., 2025; Subramonyam et al., 2024). This collaborative
role has become more evident with the recent meteoric rise in GenAI
capabilities which expanded the human-AI collaboration fields to include
complex code generation, artistic imagery, and storytelling, among
others. GenAI enabled non-experts to co-create artifacts beyond their
skill sets through natural language and gesture-based controls (Mustafa
et al., 2024; Anonymized, 2024). This synergy, pairing human strengths
(contextual awareness, judgment, and vision) with AI's computational
power, has given rise to human-AI hybrid work, where complementarity
drives shared goals (Agrawal et al., 2024; Fan et al., 2024; Järvelä et
al., 2025; Anonymized., 2024; Subramonyam et al., 2024).

On the other end of the spectrum of human-AI collaboration, AI assumes
the leading role as an independent agent that takes over the whole task
and acts as an autonomous system, e.g., decision-making algorithms or
driverless cars (Akata et al., 2020). The human role varies here from AI
being a sovereign unsupervised actor that has total control over the
task such as in autonomous trading algorithms. Here, humans may also be
passive observers who monitor the process as AI acts: humans may react
if necessary, given that total blind AI agency has shown to be neither
safe nor ethical (algorithmic bias in hiring or fatal crashes of
driverless cars) (Crandall et al., 2018; Pedreschi et al., 2025). To
mitigate total agency risks, human oversight may be embedded as a safety
mechanism, auditing or co-steering the processes in real time. This
hybrid model ensures AI's autonomy remains tethered to and constrained
by human-defined ethical frameworks and aligned with human priorities
while preserving the efficiency of agentic systems (Crandall et al.,
2018).

Beyond decisions and tasks, empathic AI has also been used across
several fields and products. Typically, AI is used to identify and
simulate emotions in human dialogue by acting as a ``friend'' or an
emotional companion. In doing so, AI may offer personalized support that
simulates some human qualities. Of course, it is not hard to imagine how
empathic AI may be useful in some situations but also, or the dangers
that can be associated with relying on AI for emotional interactions.
Already, empathic AI has a sheer volume of questionable practices and
ethical concerns, including the authenticity of its emotional responses,
manipulation, and the risk of dependency on technology for emotional
support (Akata et al., 2020).

Lately, the latest progress in AI conversation abilities has led to the
prominence of prompts. Prompts, as the name implies, indicate an
instruction that the Large Language Model (LLM) needs to act upon or
perform (Subramonyam et al., 2024; Y. Wang et al., 2024). Prompts can
also convey contextual information (background information required to
complete the task), input data (the specific question, task, or problem
to be solved), or indication of the output type (the format in which the
response should be delivered). Of course, several other types of prompts
and use cases have emerged and others may also be developed in the
future. While the concept of prompt engineering has emerged, to refine
the structuring of prompts and optimize the accuracy and relevance of
LLM outputs, it is still loosely defined and is losing relevance over
time (Gao et al., 2024).

\subsection{Review of related
literature}\label{review-of-related-literature}

In educational research, educators emphasize the role of learners as
active collaborators rather than passive recipients and AI is not an
exception (Kim et al., 2022). Kim et al.~(2022) proposed a student-AI
collaboration model based on the theory of \emph{Distributed Cognition}
that includes not only the individual but also the wider learning
settings divided into the curriculum, student-AI interactions,
environment, and evolution over time. In this conceptualization, AI is
another learning agent, rather than a learning tool, a subject of
interactions with both teachers and students, while teachers are
responsible for facilitating and designing student-AI interactions. Wang
et al.~(2021) suggested using the \emph{Theory of Mind}, which refers to
a cognitive ability to understand the mental states of others based on
behavioral and verbal cues, as a basis for modeling human-AI
interactions. Applying the Theory of Mind to human-human interactions
facilitates the building of shared expectations of others through
behavioral feedback, which in turn helps ensure cohesion and
constructive exchanges. Modeling human-AI interactions through this lens
assumes that AI should learn about and adjust students' perceptions and
expectations about AI's capabilities through behavioral cues. Currently,
learners need to adjust their mental model of an AI agent through trial
and error interactions to achieve the desired output. Coming from the
field of Human-Computer Interaction, Lodge et al.~(2023) created a
typology of human-AI interactions in education depicted as spectrums on
two axes. On the first axis, cognitive offloading, i.e.~the use of tools
to reduce the cognitive load of a task is contrasted with the extended
mind hypothesis, where tools can be used to extend individual human
cognition and function as a part of the cognitive process. On the other
axis, AI is positioned as a co-regulator of learning; however, this
approach relies on regulation skills and initiative of human learners.
At the other end of this axis are hybrid learning approaches AI that can
potentially support the social aspects of the learning process and human
cognition through personalized learning and supporting metacognitive
awareness.

Early research highlights the complex nature of student-GenAI
interactions. For example, Kim et al.~(2024) investigated student-GenAI
interactions in a collaborative public advertisement drawing task. In
particular, this study examined the role of attitudes towards GenAI and
skill level on task performance. The findings indicated that group
characteristics impacted how students used GenAI. At the same time, the
authors noted the central role of student-student collaboration in the
task process, and the potential of GenAI to enhance group collaboration
through scaffolding and regulation. Akçapınar \& Sidan (2024) compared
the performance of students in a programming task with and without GenAI
assistance. Although the performance increased in the GenAI condition,
the results also indicated that students tended to copy-paste incorrect
GenAI-generated responses. This study highlighted the need for critical
thinking skills to appropriately evaluate GenAI answers. Another study
in the domain of programming showed that students who were asked to
attempt to independently solve the task first and then receive GenAI
assistance performed better than students who started solving the
assignment with the help of AI (Singh et al., 2024). Kim et al.~(2025)
analysed student-GenAI interactions through an analysis of prompts used
in an GenAI-assisted academic writing task. The findings suggested that
students with high levels of AI literacy treated AI more collaboratively
through continuous promoting, using varied language in the prompt, and a
critical essay co-construction with GenAI, while students with low AI
literacy focused on obtaining answers to specific questions and did not
engage critically with GenAI-generated output but accepted it passively
resulting in lower quality and less original essays. Finally, a study by
Fan et al.~(2024) confirmed higher performance of students supported by
ChatGPT in an essay writing task. However, the authors did notice that
long-term learning outcomes and increased intrinsic motivation were not
enhanced by the use of ChatGPT. As such, they warned about the potential
of GenAI tools to trigger metacognitive laziness in students.

\subsection{Dynamics and process over
counts}\label{dynamics-and-process-over-counts}

The conversational nature of GenAI interactions commands a
methodological approach that takes advantage of the ``content'', the
interactive dialogue, and the temporal order of how the interactions
unfold to understand the full dimensions of the process (Fan et al.,
2024; Anonymized, 2023). An approach that goes beyond the flat and
static view of frequency-based analysis (i.e., counts) to capture the
wealth of dynamics (Molenaar \& Wise, 2022). It is also worth mentioning
that a single method may not capture the depth and complexity of such
interactions (Helske et al., 2024).

Therefore, We use a collection of methods which include sequence
analysis to capture the order, transition analysis to capture the
relations and dynamics as well as partial correlation networks to
understand how different variables interact with each other (Molenaar \&
Wise, 2022). Sequence analysis is a widely used method that helps
understand the order and linear relationships between events as they
happen. In education research ---and beyond--- sequence analysis has
been used to understand the sequential order of learning events and
capture students' behavioral patterns (Fan et al., 2022; Anonymized.,
2024; Matcha et al., 2019).

In this study, we introduce a novel method: Transition Network Analysis
(TNA). TNA is a novel method that captures both the relations, the
dynamics, and the temporal properties of the interactions (Saqr et al.,
2024). TNA ---based on Markov modeling--- is particularly well-suited
for modeling human-AI interactions due to the inherent alignment between
TNA and LLMs which both are based on the stochastic and short memory
principles (Zekri et al., 2024). LLMs have limited context windows which
mirror the limited memory of Markov property where each interaction
depends solely on the present with limited memory (Zekri et al., 2024).
On the other hand, human-AI interactions are inherently stochastic,
characterized by variability in both user behavior and LLM output
requiring a probabilistic method to model these stochastic nature
(Helske et al., 2024). TNA captures these properties through directed
probabilities allowing the analysis of interaction patterns, their
magnitude, and unique characteristics like repetitive loops, temporal
evolution, key events, and important structural patterns (Saqr et al.,
2024).

\section{Motivation for this study}\label{motivation-for-this-study}

Recent calls for paving the way for human-GenAI collaboration,
co-intelligence, or co-evolution can be heard widely in the educational
field and at the workplace at large (Järvelä et al., 2025; Mollick,
2024; Pedreschi et al., 2025). The argument is that working with GenAI
equates to working with a capable helper whose support leads to
outperforming anyone without GenAI and an outcome that exceeds either's
capabilities alone. Evidence is so far lacking beyond limited fields
---e.g., coding and writing-- that such collaboration is fruitful nor do
we have a full understanding of how it happens.

As we currency stand, the dynamics of human-GenAI collaboration have
been yet hardly studied nor has there been enough evidence of its
presumed outcome. We argue here that these dynamics will be hardly like
any human-human interaction and the outcome or the process thereof will
be dissimilar to previous models of human-machine interactions. This is
because human interactions with prompts differ significantly from how
human interactions Anonymized, 2024; Subramonyam et al., 2024; Y. Wang
et al., 2024).

In this study, we designed a cognitively demanding task that requires
students to generate a dataset that exhibits certain complex network
characteristics, e.g., centralized networks, networks with certain
community wiring or ask GenAI ---using text only--- to generate a
network that replicates real-life networks. In doing so, GenAI has to
craft complex relationships and the students have to verify that these
relationships exist and correct AI in text. We captured the interactions
between students and GenAI for analysis to map how the process of
students and GenAI unfolds across time. The interactions were coded
according to their type and several types of analysis were performed
that included TNA, key events identification as well as comparisons
across groups.

\section{Methods}\label{methods}

\subsection{Context and data
collection}\label{context-and-data-collection}

The data for this study was collected during a \emph{Network Science}
course at a European University, an elective course for undergraduate
and graduate students in Information Technology. In this course,
students learn about the basics of network analysis, their construction,
visualization, evaluation, and interpretation. As a part of the course
assessment, students had to complete four assignments with increasing
difficulty and complexity using an LLM to generate a network dataset
that met the assignment criteria:

\begin{itemize}
\tightlist
\item
  \emph{Assignment 1}: Use an LLM to create a sample network of your
  choice (e.g., food recipes, transport, interactions between countries,
  messages between people). The network should have at least 20 nodes
  and 50-100 edges with named nodes (has actual labels e.g., names, food
  or countries). Please make sure that the nodes are connected with
  interactions between the nodes that are appropriate for the chosen
  network.\\
\item
  \emph{Assignment 2}: Use an LLM to generate a network with 30-50
  nodes. Make sure that a single node is more central than any other
  node. Please refine and try to get the network structure right with
  labeled nodes.\\
\item
  \emph{Assignment 3}: Choose one network from the course literature
  (e.g., authorship network, country collaboration, disease spread)
  reading materials and try to generate a similar network using the LLM
  that shows similar communities, patterns of interactions, and node
  labels.\\
\item
  \emph{Assignment 4}: Generate a network that resembles the networks
  you built in the final project with similar structure, communities as
  well as node names and organization. The final project networks
  included networks that the students collected data for and built from
  scratch e.g., country collaborations, movie networks, and food
  recipes.
\end{itemize}

The students were required to record or take a screenshot of their
conversations with an LLM and attach it as a part of the assignment
submissions. Each assignment was graded from 1 to 10 based on the
suitability of the dataset generated and the generated networks.

\begin{longtable}[!t]{@{}
  >{\centering\arraybackslash}p{(\linewidth - 4\tabcolsep) * \real{0.3333}}
  >{\raggedright\arraybackslash}p{(\linewidth - 4\tabcolsep) * \real{0.3333}}
  >{\raggedright\arraybackslash}p{(\linewidth - 4\tabcolsep) * \real{0.3333}}@{}}
    \caption{Coding scheme for students' prompts with code frequencies,
    definitions, and examples.}\label{tbl-table1}\tabularnewline
    \toprule\noalign{}
    \begin{minipage}[b]{\linewidth}\centering
    Code
    \end{minipage} & \begin{minipage}[b]{\linewidth}\raggedright
    Definition
    \end{minipage} & \begin{minipage}[b]{\linewidth}\raggedright
    Prompt example
    \end{minipage} \\
    \midrule\noalign{}
    \endfirsthead
    \toprule\noalign{}
    \begin{minipage}[b]{\linewidth}\centering
    Code
    \end{minipage} & \begin{minipage}[b]{\linewidth}\raggedright
    Definition
    \end{minipage} & \begin{minipage}[b]{\linewidth}\raggedright
    Prompt example
    \end{minipage} \\
    \midrule\noalign{}
    \endhead
    \bottomrule\noalign{}
    \endlastfoot
    \textbf{Instruct} (n=249) & Prompts that directly address the AI, and
    provide general instruction (e.g., about formatting) & ``Create a Python
    code that generates and prints a network based on this data'' ``Now
    instead of H1 use actual names and stuff and elongate it to 40
    nodes'' \\
    \textbf{Context} (n=115) & Prompts providing context information about
    the network (e.g., the theme of the network, the definitions of edges
    and nodes). & ``Can you create an edge list of a network of interactions
    between 20 countries? The countries must be real countries and the
    interaction between them could be an exchange according to the
    students.`` ``Could you please create an edge list for a network of
    country collaborations, featuring 30 different countries and making the
    USA have the highest closeness centrality?'' \\
    \textbf{Specify} (n=239) & Prompts outlining network specifications
    (e.g., number of nodes or edges, the node with the highest centrality).
    & ``Can you extend the generated example up to about 70 lines in length?
    `` ``Could you add another edge from Russia to Kazakhstan? (Now the
    network satisfies my original requirements)'' ``Generate the 30-node
    network in CSV format'' \\
    \textbf{Disagree} (n=57) & Prompts that explicitly disagree with AI's
    output or performance, ask AI to correct or revise the output. & ``The
    list is still short. Please get the top 100 bands from IMDb again and
    look for the influences. Keep in mind that each band on the list HAS TO
    appear on the source column at least once.'' ``Your list only contains
    100 edges, which is not enough, please generate more'' \\
    \textbf{Agree} (n=15) & Prompts that explicitly agree with AI output. &
    ``Great job, thank you. Could you give each edge a weight as well,
    making the USA have the highest closeness centrality?'' ``This is a good
    start. Now, flesh out the remaining part of the file.'' \\
    \textbf{Request} (n=58) & Prompts using explicit polite requests or
    polite language (e.g., please, thank you, kindly). & ``Please make sure
    that the nodes are connected and there are some interactions..'' ``Thank
    you! Can you add more brands to the network?'' \\
    \textbf{Conclude} & The final event is the interaction and captures the
    download of the data, or closure of the interactions & n/a \\
\end{longtable}

The prompts were coded using a coding scheme for students' prompts
developed using a hybrid approach. Initially, we coded a sample of our
dataset using four codes (\emph{instruction}, \emph{context},
\emph{output indicator}, and \emph{input data}) as defined by Giray
(2023). Next, we adjusted the initial coding scheme based on the
patterns observed in our data and considering the context of the
learning activities (see Table~\ref{tbl-table1}). This included
redefining the existing codes for \emph{instruct}, \emph{context}, and
\emph{specify}, as well as adding three new codes: \emph{disagree},
\emph{agree}, and \emph{request}. One prompt could be coded with
multiple codes. The prompts were coded for the occurrence of a code
based on their order in the prompt. The coding scheme was validated by
two researchers coding a sample of student prompts (n=105). As the
coding indicated a Cohen's Kappa reliability of least substantial agree
for every code (\(\kappa\)\textgreater0.60), the remaining prompts were
coded by one researcher.

\subsection{Data analysis}\label{data-analysis}

\subsubsection{Sequence analysis}\label{sequence-analysis}

The analysis of the data was performed using a collection of methods to
capture the multifaceted temporal dimensions of human-AI interaction.
First, sequence analysis was used to study the temporal unfolding of
students' prompts throughout a full conversation with the LLM. Sequence
analysis is a well-established method in educational research commonly
used to examine students' sequences of actions or events in learning
contexts (Fan et al., 2022; Matcha et al., 2019).

In our study, students' coded prompts were temporally ordered and
prepared for sequence analysis where the codes identified in the prompts
(see Table~\ref{tbl-table1}) were the alphabet. The time epoch was the
full conversation with the LLM of each assignment. As such, a full
sequence was created for each conversation. The sequences of all
students' conversations were used to build a sequence object using the
\emph{TraMineR} package (Gabadinho et al., 2011). A sequence index plot
was generated to visualize the linear temporal progression of the coded
prompt sequences, providing a clear representation of interaction
patterns over time. We also created sequence plots to compare high and
low achievers and to compare early and late assignments using the same
procedure.

\subsubsection{Transition network
analysis}\label{transition-network-analysis}

To capture the temporal and relational dynamics of the interaction
process we used TNA (Saqr et al., 2024). TNA offers a robust
theory-informed framework for capturing the dependencies and
relationships between events. The visualization of TNA offers a bird's
eye view of interactions. Furthermore, TNA helps identify key events
(centralities) and transitions (edge strengths). Moreover, TNA
incorporates techniques like bootstrapping to validate the significance
of the modeled process to ensure that the insights are robust and not
artifacts of the dataset. All of such features make TNA optimal for
capturing the process of interactions over ``static'' methods such as
social network analysis or process mining (López-Pernas et al., 2024;
Saqr et al., 2024).

TNA models the dynamics of transitions between coded prompts based on
Markov Modeling (Helske et al., 2024; Saqr et al., 2024). The TNA
network, \emph{G=(V, E, W)}, is defined by: nodes \emph{V} (coded
prompts), edges \emph{E} (directed transitions between coded prompts),
and weights \emph{W} (transition probabilities). Directed edges \emph{E}
reflect the direction of the transitions and weights (transition
probabilities) represent the probability of transitioning from the
source prompt to the target prompt. Weights quantify the likelihood of a
prompt following another, thus indicating the magnitude, direction, and
relationships of prompts. The analysis of the TNA network helps identify
the prompt sequences, key transitions between prompts, and central
prompts acting as interaction hubs. Transition patterns show the
recurring combinations of events observed throughout the interaction
process and represent consistent behaviors that characterize a learner's
typical approach to interacting with AI.

In this study, we used the R package \emph{tna} (López-Pernas et al.,
2024) to estimate the TNA network from the sequence object of coded
interactions described in the previous section, to model the structure
and dynamics of student-AI interactions. To find the key transition
events we computed the in-strength centrality as the sum of inbound
transitions to each node to capture the importance of the node as a
target of transitions (Newman, 2018). We also estimated the node
betweenness centrality to capture the node's importance in mediating
other prompts and bridging different types of interactions. Edge
betweenness centrality shows the critical transitions that mediated
students-AI interaction and therefore were estimated and plotted (Bae \&
Kim, 2014).

We also estimated separate TNA networks for low achievers and high
achievers. For comparison between these two TNA networks, we used a
resampling-based permutation test. TNA permutation test calculates the
statistically significant edges and centrality measures by permuting the
input data (1000 times) and generating a null distribution of either
network under the null hypothesis of no difference. P-values are then
computed based on the proportion of permuted differences that are as
extreme as the observed ones (van Borkulo et al., 2022). The permutation
test was used to compare early and late assignments, high and low
achievers as well as short and long sequence networks.

To rigorously evaluate the magnitude of difference between the
processes, we computed the effect size, the Standardized Mean Difference
for permutation (SMD\_perm) by determining the observed difference
between the means of two groups and then dividing it by the standard
deviation, resulting in the SMDperm value. A value near zero suggests
that the observed difference is within the range of random fluctuation,
whereas an absolute value greater than one indicates a moderate effect.
Values with an absolute magnitude exceeding two point to a strong
effect, and those greater than three denote a very large effect that is
highly unlikely under the null hypothesis (Andrade, 2020).

To compare the frequency of interactions between high and low achievers,
we employed a chi-squared test alongside a mosaic plot with
Pearson-residual shading. The chi-square test provides a formal
statistical evaluation of whether the observed frequencies deviate
significantly from those expected under the null hypothesis. We also
used a mosaic plot to offer a visual representation of how the
interactions differ across the examined groups (Meyer et al., 2007). In
the mosaic plot, the size of each tile reflects the proportion of cases
in a given category, while the shading within these tiles indicates both
the magnitude and direction of the differences between observed and
expected frequencies. More intense shading signifies a higher likelihood
of deviation, with blue tones representing categories that are more
likely than expected, and red tones indicating categories that are less
likely (Meyer et al., 2007).

\subsubsection{Psychological networks}\label{psychological-networks}

Lastly, we use psychological networks to understand the relationships
between the events and students' grades. Psychological networks are
adept at capturing the complex interactions and dependence between
different variables (Epskamp, Borsboom, et al., 2018). ---in our case,
performance and prompt types. In doing so, this method overcomes the
linear assumptions of linear models that are hard to satisfy in a
complex process where each action depends on the other, as is the case
in the interactions between humans and AI (Epskamp, Waldorp, et al.,
2018).

We estimated a regularized partial correlation network using the R
package \emph{bootnet} (Epskamp, Borsboom, et al., 2018), where nodes
represent variables (frequency of codes that each student has used as
well as their grades). The edges represent the regularized partial
correlations between nodes after controlling for others (``ceteris
paribus''); no edge means conditional independence. Regularization
reduces spurious and trivial edges and eliminates small negligible
correlations resulting in simpler, sparser, and more interpretable
networks. Networks are usually undirected, signed, and weighted by
partial correlation magnitude.

\section{Results}\label{results}

\subsection{Descriptive statistics}\label{descriptive-statistics}

Our dataset contains 122 conversations by 49 students, with a total of
289 prompts and a total of 726 interactions. Each conversation had 5.95
prompts on average with a median of 4 and Standard Deviation of 5.88.
Prompts ranged from 2 to 47 in length. The most frequent code was
\emph{Instruct} (see Table \#tbl-table2), which appeared 249 times,
being present in 88.61\% of the prompts and at least once in each
conversation. Closely following was \emph{Specify} (n = 237), which
appeared in 84.34\% of the prompts, and in almost all conversations (n =
119, 97.54\%). \emph{Context} was next, appearing 110 times, in 39.15\%
of the prompts and 101 conversations (82.79\%). \emph{Disagree} was far
less common (n = 59), only present in 21\% of prompts, 20.49\% of the
conversations. Lastly, \emph{Agree} was the least common type of
interaction (n = 13), only appearing in 4.63\% of all prompts, 5.74\% of
all conversations.

\begin{longtable}[]{@{}
  >{\raggedright\arraybackslash}p{(\linewidth - 8\tabcolsep) * \real{0.2000}}
  >{\raggedleft\arraybackslash}p{(\linewidth - 8\tabcolsep) * \real{0.2000}}
  >{\raggedleft\arraybackslash}p{(\linewidth - 8\tabcolsep) * \real{0.2000}}
  >{\raggedleft\arraybackslash}p{(\linewidth - 8\tabcolsep) * \real{0.2000}}
  >{\raggedleft\arraybackslash}p{(\linewidth - 8\tabcolsep) * \real{0.2000}}@{}}
\caption{Descriptive statistics of the coded
prompts}\label{tbl-table2}\tabularnewline
\toprule\noalign{}
\begin{minipage}[b]{\linewidth}\raggedright
Code
\end{minipage} & \begin{minipage}[b]{\linewidth}\raggedleft
Frequency
\end{minipage} & \begin{minipage}[b]{\linewidth}\raggedleft
\% of prompts
\end{minipage} & \begin{minipage}[b]{\linewidth}\raggedleft
Number of conversations
\end{minipage} & \begin{minipage}[b]{\linewidth}\raggedleft
\% of conversations
\end{minipage} \\
\midrule\noalign{}
\endfirsthead
\toprule\noalign{}
\begin{minipage}[b]{\linewidth}\raggedright
Code
\end{minipage} & \begin{minipage}[b]{\linewidth}\raggedleft
Frequency
\end{minipage} & \begin{minipage}[b]{\linewidth}\raggedleft
\% of prompts
\end{minipage} & \begin{minipage}[b]{\linewidth}\raggedleft
Number of conversations
\end{minipage} & \begin{minipage}[b]{\linewidth}\raggedleft
\% of conversations
\end{minipage} \\
\midrule\noalign{}
\endhead
\bottomrule\noalign{}
\endlastfoot
Instruct & 249 & 88.61\% & 122 & 100.00\% \\
Specify & 237 & 84.34\% & 119 & 97.54\% \\
Context & 110 & 39.15\% & 101 & 82.79\% \\
Disagree & 59 & 21.00\% & 25 & 20.49\% \\
Request & 58 & 20.64\% & 43 & 35.25\% \\
Agree & 13 & 4.63\% & 7 & 5.74\% \\
\end{longtable}

\subsection{The temporal unfolding of the
interactions}\label{the-temporal-unfolding-of-the-interactions}

\begin{figure}

\centering{

\includegraphics[width=0.8\textwidth]{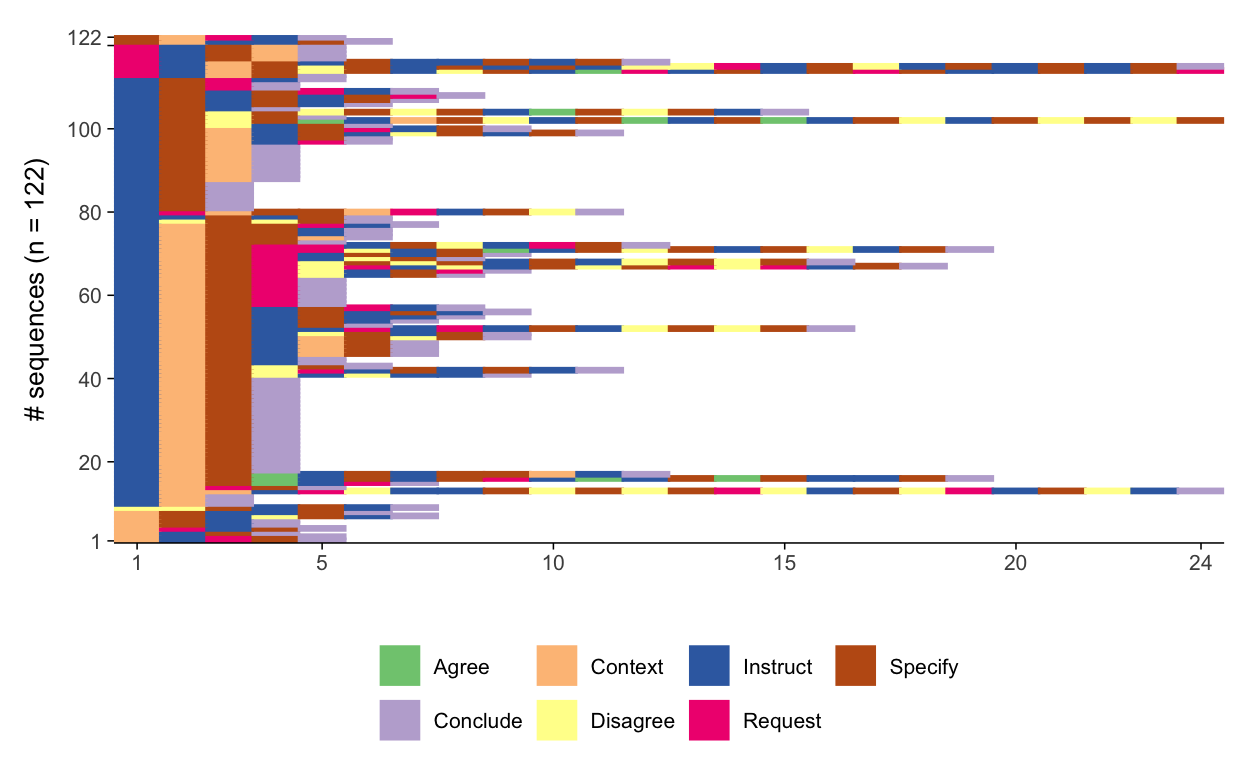}

}

\caption{\label{fig-image1}Sequence index plot of students' coded
prompts}

\end{figure}

To map the linear sequential unfolding of interactions we used sequence
analysis. A sequence index plot was plotted in Figure~\ref{fig-image1}
to show the sequence of codes in each conversation. Each horizontal line
in the figure represents a single conversation where the colored blocks
are the codes in the order they were identified in the prompts.

The most common starting point is \emph{Instruct} (n = 103, 84.43\% of
the conversations), and with far less frequency \emph{Request} or
\emph{Context} (n = 8, 6.56\%), and almost none of the remaining codes.
The most frequent starting sequence was \emph{Instruct} - \emph{Context}
- \emph{Specify} (n = 53, 51.64\%)\emph{,} which in a moderate
percentage of the conversations was enough to \emph{Conclude} (n = 23,
18.85\%), although it was often followed by \emph{Request} (n = 15,
12.29\%), or with further iterations of \emph{Instruct} (n = 14,
11.48\%), or \emph{Agree}/\emph{Disagree} (n = 3, 2.45\%). The
overarching picture is instruction, refining with contextual details and
specifications and in some cases, a long process where students struggle
with finding the solution that extends over long sequences.

\begin{figure}

\centering{

\includegraphics[width=0.8\textwidth]{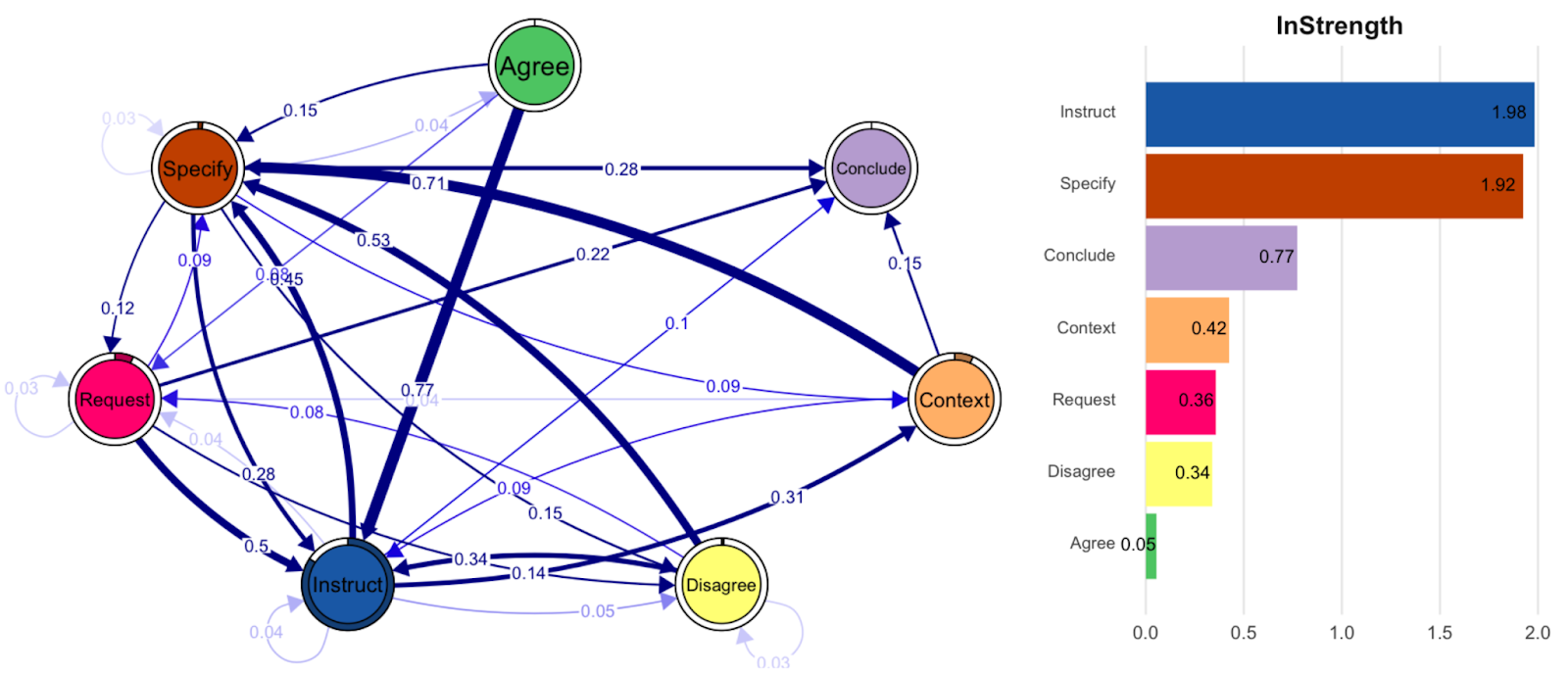}

}

\caption{\label{fig-image2}Circles represent coded interaction prompts,
the pie around each pie represents the initial probability (the
proportion of times students started with this interaction, the edges
are directed arrows where the numbers are the transition probabilities
between interactions.}

\end{figure}

Beyond sequences, transitions can help show how interactions unfolded
and what the structure of the process looks like. In
Figure~\ref{fig-image2} we plot the transitions between codes within the
same conversation. Circles (Nodes) represent coded interaction prompts.
Pie around each indicates the initial probability, which is the
proportion of times students started with that particular interaction.
Directed Arrows (Edges) show the transition probabilities between
different interactions. The numbers on the arrows represent the
likelihood of transitioning from one interaction to another.

The TNA model illustrates a dynamic process where \emph{Instruct} is the
primary central event, starting with a high initial probability of
84.43\% and receiving the largest total number of incoming transitions
from other states (totaling approximately 1.98). Significant sources of
transitions originate from \emph{Agree}, with 76.9\% of transitions from
\emph{Agree} leading to \emph{Instruct} indicating approval and
instruction for more in the same direction. This is followed by
transitions from \emph{Request} (50.0\%, reflecting appreciation and
emotional engagement with the LLM) and \emph{Disagree} (33.9\%,
reflecting need for corrections. It is worth noting that agreement is
followed by more instruction, disagreement is followed by more specific
prompts.

From \emph{Instruct}, students frequently transition to \emph{Context}
(31.3\%), indicating a refinement of their prompts to provide more
specific contextual details. \emph{Specify} also stands out as a key
event, receiving a substantial number of incoming transitions (totaling
approximately 1.92) from various states. Notable sources include
\emph{Context} (70.9\%), \emph{Instruct} (45.0\%), and \emph{Disagree}
(52.5\%). The transitions from \emph{Disagree} to \emph{Specify}
(52.5\%) and \emph{Instruct} (33.9\%) highlight a student strategy of
iterative correction. This suggests that rather than abandoning the
task, students persist by refining prompts or restarting instructions,
even when experiencing disagreement or frustration. The transition from
\emph{Request} to \emph{Instruct} (50.0\%) further underscores the
courteous nature of student interactions with the LLM, even while giving
instructions. \emph{Conclude} functions as the terminal endpoint in this
TNA model. \emph{Conclude} receives the highest proportion of
transitions from \emph{Specify} (28.3\%), highlighting the role of
detailed prompt refinements in reaching task closure. Furthermore,
transitions from \emph{Request} to \emph{Conclude} (22.4\%) suggest that
students often express appreciation before finalizing their
interactions. Transitions from \emph{Context} (15.5\%) and
\emph{Instruct} (9.6\%) to \emph{Conclude} further emphasize the
importance of contextual framing and initial guidance in reaching the
final \emph{Conclude} state of the interaction. The overarching dynamics
paints a picture of trial, re-trial and gauging or guessing in an
iterative process of refinement dominated by instructions and refined
instructions that ultimately lead to conclusion of the task.

\begin{figure}

\centering{

\includegraphics[width=0.8\textwidth]{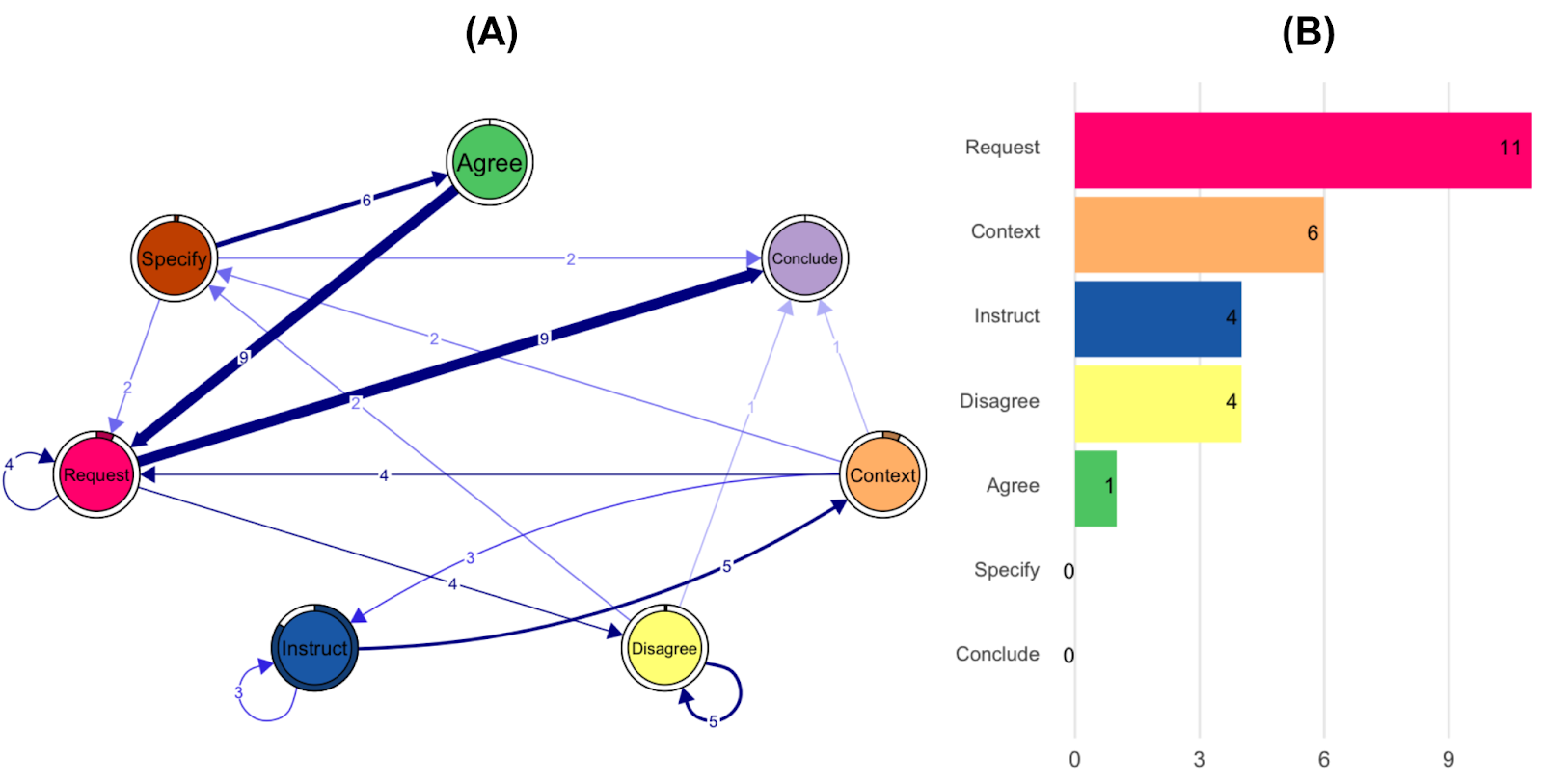}

}

\caption{\label{fig-image3}(A) Edge betweenness centrality. (B). Node
betweenness centrality}

\end{figure}

To understand which transitions were the most central, we computed the
node and edge centralities of betweenness (Figure~\ref{fig-image3}). The
node betweenness centrality showed \emph{Request} and \emph{Context}
mediated other interactions more than other prompt types. Looking closer
at the network of edge betweenness to capture the critical transitions,
we can see a pattern of flow starting from \emph{Specify}, \emph{Agree}
to \emph{Request} and later concluding the interaction. Such a pattern
emphasizes the value of refining based on reaching an agreement between
what the student wants and what the LLM provides which mediates a
successful completion of the task.

\subsection{Assignment complexity}\label{assignment-complexity}

The correlation between assignment complexity---as indicated by the
order in which assignments, increasing in complexity, were
presented---and grades was minimal, r(120) = .081, p = .40, with a 95\%
confidence interval ranging from --0.10 to 0.26. In other words, a
non-statistically significant relationship was observed among assignment
complexity, prompt length, and students' grades. Similarly, the length
of students' prompts was not significantly correlated with their grades,
r(120) = .036, p = .70, 95\% CI {[}--.14, .21{]}. These results indicate
that students' conversations were neither particularly longer in complex
problems, nor was it longer or shorter in high achieving students.

\begin{figure}

\centering{

\includegraphics[width=0.8\textwidth]{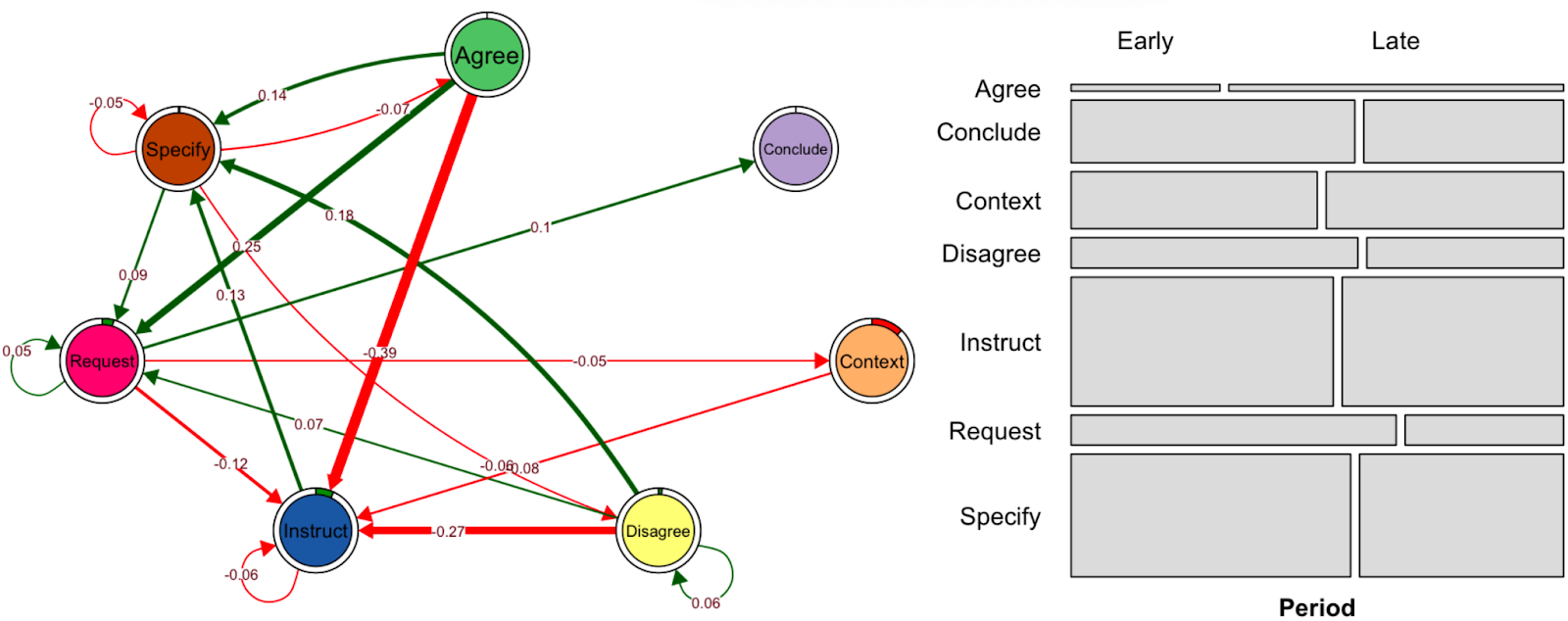}

}

\caption{\label{fig-image4}Left: TNA network representing the difference
between the first two (early) and last two (late) assignments. Right:
Mosaic plot representing the chi-squared test comparing the frequency of
each prompt code between early and late assignments.}

\end{figure}

Beyond counts, to investigate whether complex problems would result in
different dynamics, we estimated two networks for early and late
assignments (early assignments are simpler and late assignments are more
complex). Both assignments were compared using a permutation test to
find out the significant transitions and then plotted (see
Figure~\ref{fig-image4} Left). Furthermore, a Pearson's Chi-squared test
was used to compare the differences in frequencies along with a Mosaic
plot to further offer a granular picture of the significant prompts.

The overall picture of early versus late dynamics show more transitions
towards \emph{Specify and Request} less towards \emph{Instruct} and
slight decrease in \emph{Context}. However, the statistically
significant transitions were from \emph{Instruct} to \emph{Specify},
\emph{Specify} to \emph{Request} and \emph{From} \emph{Request} to
\emph{Context} as well as two loops (indicating repetitions) in Instruct
and Specify, both had the highest effect size SMD\_perm of 2.4 and 2.3
respectively. This repetitive pattern indicates an inability to reach a
closure and continuing to repeat their requests again and again
Table~\ref{tbl-table3}.

The mosaic plot visualizes the distribution of categories across Early
and Late Assignments (see Figure~\ref{fig-image4} Right). The Chi-square
test results (\(\chi\)² = 9, df = 6, p = 0.2) indicate no significant
association. This suggests that prompt distributions remain relatively
invariant over time. An overall picture that can be summarized as more
asking ``please do this with these specifications'' that lingers on.

\begin{longtable}[]{@{}cccc@{}}
\caption{Statistically significant differences in transition
probabilities (edges)}\label{tbl-table3}\tabularnewline
\toprule\noalign{}
Edge name & Difference & Effect size & p-value \\
\midrule\noalign{}
\endfirsthead
\toprule\noalign{}
Edge name & Difference & Effect size & p-value \\
\midrule\noalign{}
\endhead
\bottomrule\noalign{}
\endlastfoot
Request -\textgreater{} Context & -0.05 & -1.5 & 0.04* \\
Instruct -\textgreater{} Instruct & -0.06 & -2.4 & 0.01* \\
Specify -\textgreater{} Request & 0.09 & 2.0 & 0.049* \\
Instruct -\textgreater{} Specify & 0.13 & 1.9 & 0.04* \\
Specify -\textgreater{} Specify & -0.05 & -2.3 & 0.02* \\
\end{longtable}

\subsection{Conversation length}\label{conversation-length}

\begin{figure}[!ht]

\centering
\includegraphics[width=0.7\linewidth]{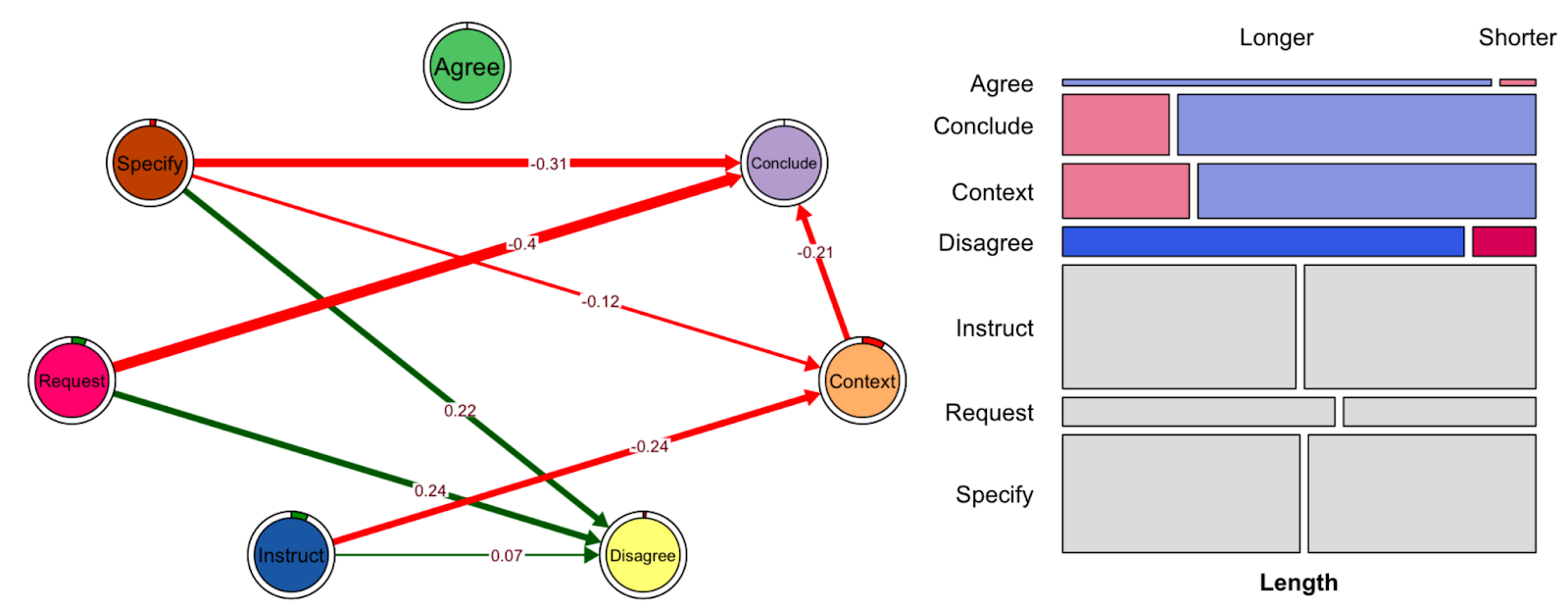}

\caption{\label{fig-image5}Left: TNA network representing the difference
between long and short conversations. Right: Mosaic plot representing
the chi-squared test comparing the frequency of each prompt code between
long and short conversations.}

\end{figure}

To take a deeper look at the long sequences ---upper quartile---
sequences exceeding the 75th percentile in which students took rather
longer conversations with the LLM, we estimated TNA networks for longer
and shorter prompt sequences. A permutation test was performed to find
out the significant transitions (Figure~\ref{fig-image5} Left) and a
Pearson's Chi-squared test was performed to compare the distribution of
frequencies of longer versus shorter sequences with a Mosaic plot
(Figure~\ref{fig-image5} Right). The results show more transitions
towards \emph{Disagree}, less to \emph{Conclude} and \emph{Context}
which is a collective picture of frustration rather than asking for more
precise output, students were more expressive of disapproval and asking
for the LLM to re-do in a different way.

\begin{figure}

\centering{

\includegraphics[width=0.8\textwidth]{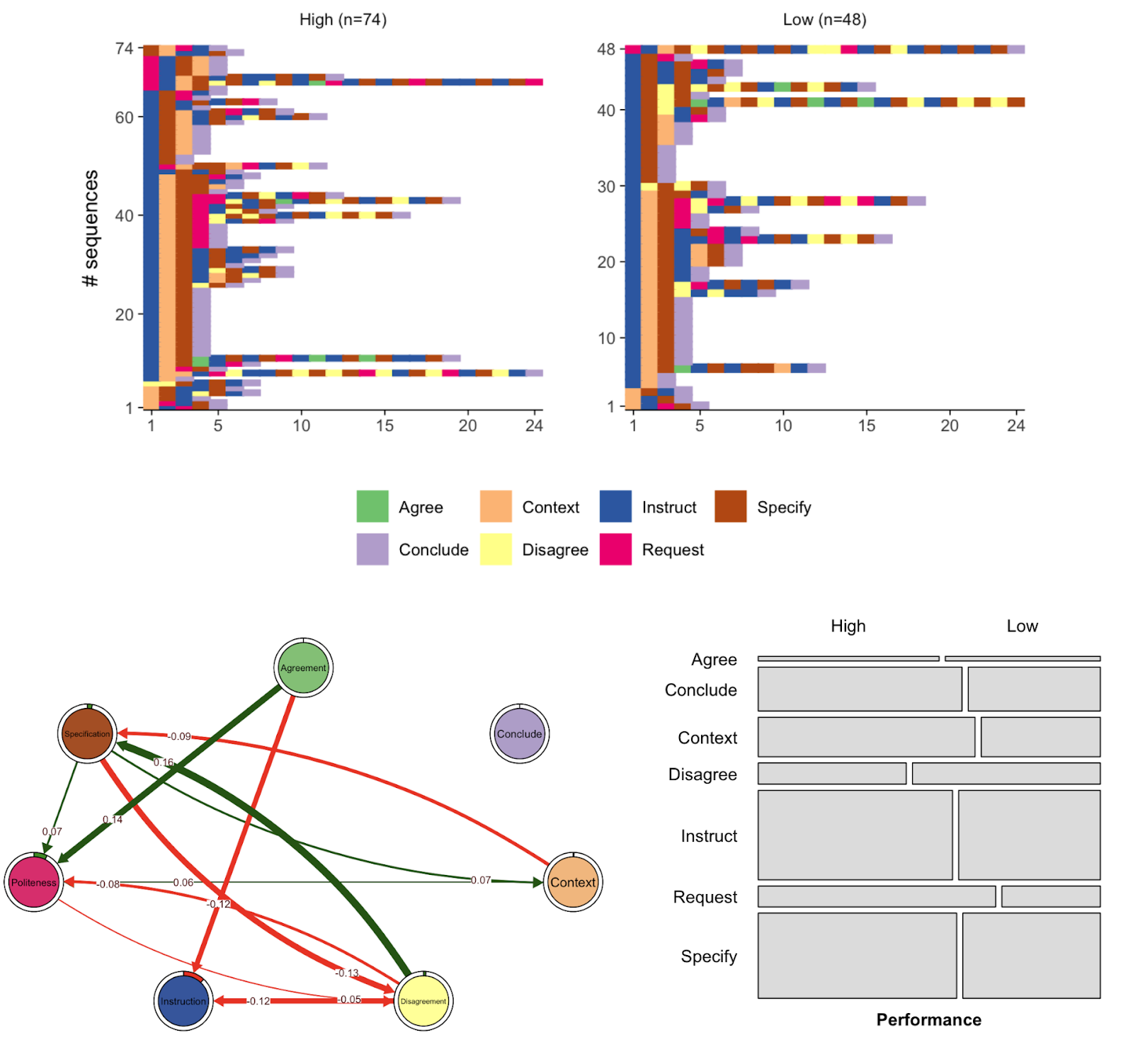}

}

\caption{\label{fig-image6}Top: Sequence index plot for high performers
(left) and low performers (right). Bottom left: TNA network of the
difference between high and low performers. Bottom right: Mosaic plot
representing the chi-squared test comparing the frequency of codes
between high and low performers.}

\end{figure}%

\subsection{Relationship with performance}\label{relationship-with-performance}

To examine if interaction with the LLM was a function of achievement
where high performing students have a different ---or maybe better
approach--- compared to the low performing students. We estimated two
TNA networks and computed a permutation test to find out the statistical
significance of the differences in transitions (Figure~\ref{fig-image6}
bottom left). We also created a sequence index plot for each group
(Figure~\ref{fig-image6} top) and compared the distribution of prompts
with a chi-squared test and a mosaic plot (Figure~\ref{fig-image6}
bottom right). Overall, the differences were subtle and statistically
insignificant sequences in every test.

\begin{figure}

\centering{

\includegraphics[width=0.8\textwidth]{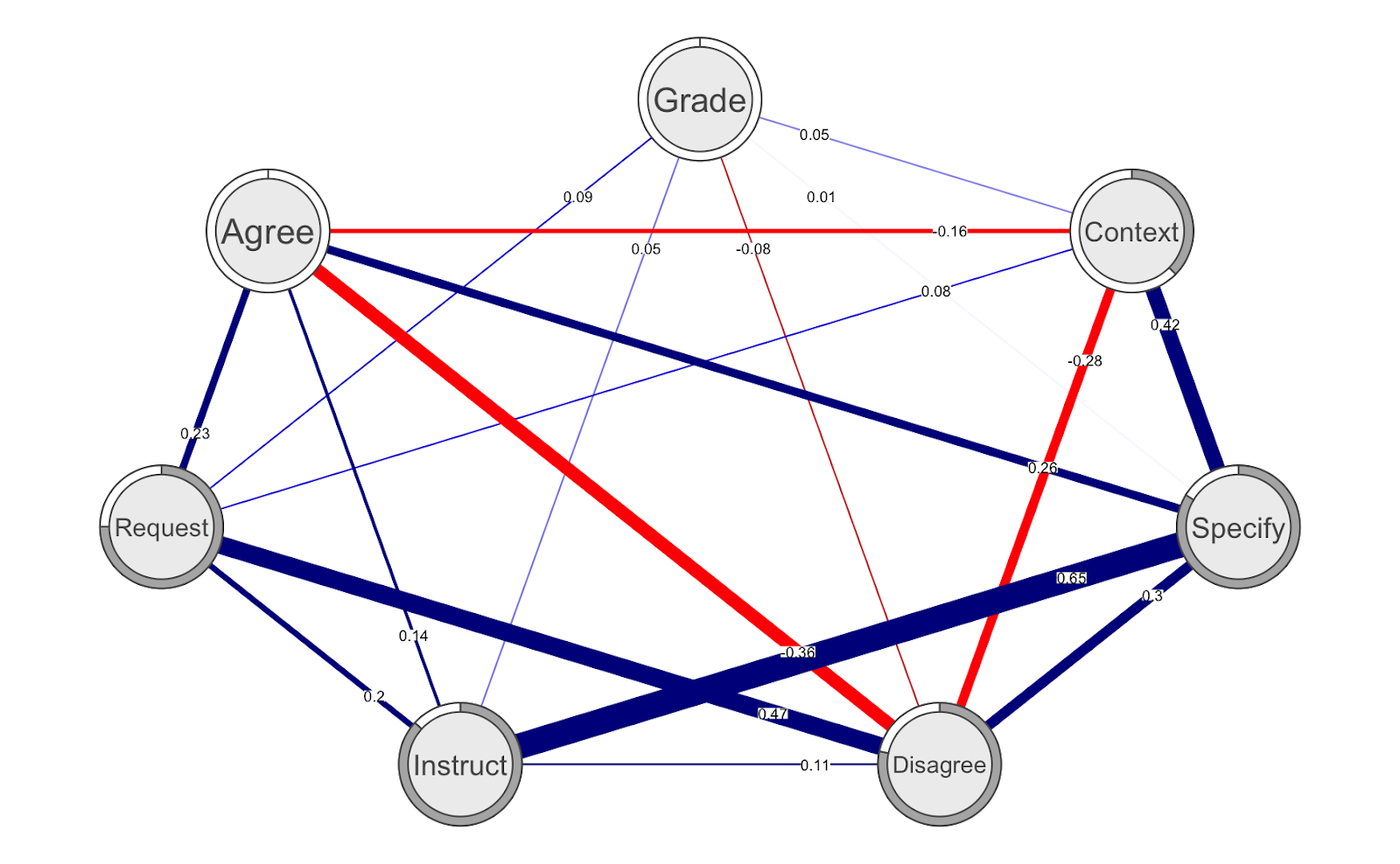}

}

\caption{\label{fig-image7}Regularized partial correlation network
between the assignment grade and the frequency of each code}

\end{figure}

Despite being insignificant, a descriptive analysis may shed light on
some differences (Figure~\ref{fig-image7}). The most notable difference
in sequences was that high achievers tend to start with \emph{Request}
(+0.07) before \emph{Instruct} (-0.119). At time point 2, high achievers
use more \emph{Instruct} (+0.06) and \emph{Context} (0.05) but less
\emph{Specify} (-0.132). At time point 3, high achievers use less
\emph{Conclude} less (-0.11), more context (+0.085) and continue to have
slightly higher \emph{Context} at time point 4 with (+0.056) and
\emph{Request} (+0.057). The TNA networks also show subtle differences.
The highest differences can be seen in the transition between
\emph{Disagree} and \emph{Specify} where high achievers refine their
prompts (+0.16), \emph{Specify} to \emph{Request} (0.07) and from Agree
to Request (+0.14). We also see a mutual transition between
\emph{Specify} and \emph{Request} (+0.07 and +0.04) and from
\emph{Specify} to \emph{Context} (+0.07). All of such transitions show a
picture of successful handling of the LLM and steering the conversation
for successful completion with requesting language and more
specifications. However, a permutation test comparing the transition
across groups showed no statistical significance difference between both
high and low achievers.

In the same vein, a regularized partial correlation network between the
grade of the assignment and the frequency of each interaction type was
plotted to understand the complex interplay between all the involved
factors and more importantly to show the correlations of interactions
and grades. The network shows a positive correlation between
\emph{Request} and grades (\emph{pcor}=0.09) after controlling for all
other variables in the network. We also see a positive correlation of
grades with \emph{Instruct} (\emph{pcor}=0.05) and \emph{Context}
(\emph{pcor}=0.09). Higher grades were less associated with
\emph{Disagree} (\emph{pcor}=0-0.09). The R2 of grades were a very small
indication that grades cannot be explained by the type of interactions.

\section{Discussion}\label{discussion}

The emergence of new GenAI technologies is widely hyped to reshape
education, technologies and how humans use or interact with education
technology at large (Järvelä et al., 2025; Pedreschi et al., 2025).
While human-technology interactions have been the center of science for
a long time, the new realities necessitates an investigation of how
humans interact with GenAI especially in dealing with complex problems,
a phenomenon that our study aimed to investigate.

Notwithstanding the conversational capabilities of GenAI that can be
felt as human, or the high performance on cognitive testing benchmarks,
the overall picture in our study has been ``instruct, serve, repeat'',
where the students are trying to translate the problem to instructions
and the AI is trying to understand and respond as far it can get the
actual intentions in terms of what it can actually perform (Subramonyam
et al., 2024). In particular, students used mostly short prompts of
repetitive instructions, requesting, asking for more repeating ``this
part was good'', ``add more here'', ``not this way'' ``please''. The
whole dynamic was marred by a distinct lack of cognitive alignment
between the interacting sides: human and LLM. Often, students are caught
in a process that extends over long sequences where they struggle to
find a solution. A process of trial, re-trial and gauging or guessing,
and requesting a repeat. A process that highlights the underlying
disconnect ---the lack of a shared framework or common language. In
other cases, students opted for a one- or two-shot interaction with the
LLM, accepting the LLM's first output without further refinement. These
instances may be an indication of students' overreliance on AI-generated
responses.

Prompting or instructing LLMs to generate outcomes thus becomes shallow,
where the student tries some heuristics and the LLM generates immediate,
compliant responses based on direct user instructions, leaving little
room for deep cognitive engagement (Bachaalany, 2024; Y. Wang et al.,
2024). A loop ensues where the users and LLM try to guess each other's
intentions, bypassing iterative, reflective processes essential for
critical thinking, problem-solving, and deeper inquiry. In doing so, the
process reduces the opportunities for dialogue, exploration of
alternative perspectives, or metacognitive refinement of their own
reasoning (Fan et al., 2024; Stadler et al., 2024). This rigid
interaction model risks reinforcing surface-level understanding, where
users passively accept AI-generated content rather than engaging in
meaningful cognitive effort to evaluate, question, and refine their
inquiries (Subramonyam et al., 2024; Y. Wang et al., 2024). The quality
of GenAI-student exchanges may also rely on student's AI literacy, which
is a skill that encompasses both productive communication with AI, but
also a critical engagement with its output (Kim et al., 2025).

A study by Marrone et al.~(2024) on student perceptions on AI as a
collaborator in solving complex problems indicated that students can
have unrealistic expectations of the AI capabilities. Current LLMs
cannot recognize and react to the users misconceptions about their
capabilities to guide the problem-solving process, which leads to poor
and unproductive exchanges (Wang et al., 2021). While the iterative
instruction may eventually lead to a conclusion, it is definitely not
what educators aspire to have in a cognitively stimulating academic
environment. As shown by Singh et al.~(2024) the immediate involvement
of AI to solve a task may actually lead to worse results than attempting
to solve a problem without assistance and use AI to refine the solution.

Furthermore, the lack of any correlation with complexity of assignment
or achievement on any measure be it the counts, the sequences or the
process emphasizes several conclusions. First, LLMs may be leveling or
removing barriers from complex tasks. This raises a serious challenge
for assessment and how teachers approach learning tasks. Second, the
lack of consistency in LLM's responses shows a fundamental issue: they
are good frequently, but unpredictable. This led to excessively random
long conversations where both the student and the LLM were clueless of
how to conclude the task. The lack of correlation with the achievement
further emphasized this finding that AI may obviate the borders between
high and low achievers, but also, it may make both frustrated in some
tasks. Taken together, our findings paint a picture of a process
dominated by interactions between different mental models and limited
mutual understanding, rather than fostering cognitive engagement that
advances the students' thinking.

Taken together, our study shows that the existing LLMs are not ideal for
the hypothesized human-AI collaboration. By design, LLMs accept
interactions through ``prompts'' and best models to respond to prompts
are those designed as Instruction-tuned LLMs. One can say that these
models are ---by design--- followers, not partners. As such, the burden
of co-regulation of the learning process still relies heavily on student
individual characteristics and initiative (Kim et al., 2024; Lodge et
al., 2023). It leads to a question, to what extent a LLM can be treated
as an equal collaborator rather than a tool used to solve a problem. In
fact, published system prompts from major AI companies often stress the
importance of user satisfaction and compliance as primary objectives in
a manner that maximizes obedience and minimizes argumentation or
disagreement (Bachaalany, 2024). While we have tried to use a complex
task in our study, it made little difference given that the other
partner in collaboration is instructed to avoid a cognitively engaged
collaboration that educationists should pursue.

Interactions with LLMs present a unique set of cognitive challenges
which stem from the differences in how humans and LLMs process
information. When humans are faced with a problem, they first form
abstract intentions and goals which they try to solve through an
iterative multistep process that involves feedback and evaluation (Gao
et al., 2024; Li et al., 2024; Y. Wang et al., 2024). The process
entails refinement, negative feedback and complex interplay of mental
processes. While with LLMs, humans try to convert their goals to
``flat'' written language (instructions) devoid of real-world physical
underpinnings and the LLMs engage in a routine of executing these
instructions based on matching with complex patterns within its training
data. This misalignment has been referred to as a blind spot or a gulf
of misunderstanding between the two partners (Subramonyam et al., 2024).

These challenges are often compounded by the opaque nature of LLM
decision-making, which hinders users from forming accurate mental models
of how responses are generated (Li et al., 2024; Y. Wang et al., 2024).
This opacity makes it difficult for users to trust or scrutinise the
LLM's output, resulting in either over-reliance on or unwarranted
dismissal of AI output (Mustafa et al., 2024; Y. Wang et al., 2024).
This opacity---whether by design or by nature---leaves users unable to
know the best way to produce the desired responses. Furthermore, the
``stochastic'' ever-changing nature of LLMs and the advances in how they
process input or produce output render learnt skills obsolete quite fast
(Crandall et al., 2018). Furthermore, the coherent and seemingly
authoritative tone of LLMs, with no explicit indicators of uncertainty,
may fool users into relying on their output without checking or engaging
in a critical evaluation. As demonstrated in the study by Akçapınar \&
Sidan (2024), where the uncritical view of the genAI output led several
students to direct copy-paste incorrect programming solutions into their
assignments.

If history has taught us anything, it is that the increasing
capabilities of GenAI have always been accompanied by a growing reliance
on AI as a machine where humans delegate more tasks to AI (Akata et al.,
2020; Crandall et al., 2018). As AI has advanced to generate
sophisticated content and creative outputs, humans have increasingly
turned to these systems for more cognitive tasks to ``do the thinking''
(Gao et al., 2024; Li et al., 2024). However, this shift toward
AI-generated solutions comes with its drawbacks (Risko \& Gilbert, 2016;
Stadler et al., 2024). Unlike traditional knowledge construction, which
involves actively searching, evaluating, and synthesizing information,
AI often delivers direct and easily accessible answers that may lack the
socio-emotional intelligence and human values embedded in the
traditional process (Stadler et al., 2024). This convenience, while
useful, risks undermining the development of critical thinking,
creativity, and problem-solving skills, leading to passive knowledge
consumption and offloading cognitive tasks to the AI tools (Risko \&
Gilbert, 2016). Risko and Gilbert (2016) describe cognitive offloading
as using external tools, such as AI, to reduce mental effort. While AI
may automate or bring efficiency to some tasks, over-reliance on AI
risks diminishing human cognitive skills and fostering dependency
leading to erosion of cognitive skills. As we have seen in our study, AI
was not the collaborator we are after, nor were students stimulated to
use it in a collaborative way. Nevertheless, it may make some tasks
easier.

To that end, educators may need to build custom AI models that are
designed to be safe, collaborative, cognitively engaging as well as
stimulating. For instance, recent research has referred to the term
``proactiveness'' in conversational agents such as LLMs, as a way to
overcome the current passive question-answer paradigm (Liao et al.,
2023). The first steps towards proactive GenAI are seen in modern LLMs
where follow-up questions are suggested for the user to continue the
conversation. Being able to train the GenAI models to adaptively curate
these follow-up questions based on sound pedagogical principles would
stimulate productive collaboration. Ideally, the GenAI would be able to
enact a scripted role such as summarizer or planner (Strijbos \&
Weinberger, 2010), in line with the literature on computer-supported
collaborative learning.

\section{Limitations}\label{limitations}

Our study has limitations concerning the coding of the prompts. Despite
our best efforts to ensure validity, the inherent subjectivity of human
coding remains a factor that cannot be entirely eliminated. The
stochastic nature of large language models (LLMs) introduces variability
in their responses across iterations. However, by employing a large
number of prompts and incorporating diverse students and tasks, we
believe we captured the behavioral patterns of LLMs. As demonstrated in
our findings, LLMs' patterns exhibited repetitive and stable
characteristics across different tasks and students, confirming that the
behavior of LLMs in this context is not random but follows a structured
pattern. Furthermore, LLMs behavior in our study can be traced primarily
to their system prompts, which are hardwired directives shaping their
responses. Consequently, our findings reflect an inherent structure in
LLM behavior that persists despite ongoing advancements in model
architecture. Beyond LLM considerations, our methodological approach
also carries limitations. TNA is constrained by the assumptions of
Markov models, which presume that present states are conditionally
independent of past ones. While this assumption may simplify the
complexities of social, temporal, and multicomponent interaction
processes, it remains a necessary abstraction to manage the complexities
of dynamic interactions. Similarly, sequence analysis, while effective
in mapping linear successions of events, does not capture other temporal
dependencies such as cyclical or overlapping interactions. Despite these
constraints, our study mitigated these methodological limitations by
integrating four different analytical approaches, ensuring a more
comprehensive understanding of LLM behavior across multiple dimensions.

\textbf{Acknowledgement}

The authors would like to express their sincere gratitude to the
participants, whose willingness to take part in this study made this
research possible.

\textbf{Availability of data and material}

The data is available upon a reasonable request from the authors and
subject to ethical approval from the governing body.



\bibliographystyle{apa}


\end{document}